\documentclass[11pt]{article}

\usepackage{latexsym}
\usepackage{graphicx}
\usepackage{amssymb,amsmath}

\def \eq#1{Eq.~(\ref{#1})}
\def \eqs#1#2{Eqs.~(\ref{#1})--(\ref{#2})}

\def \fig#1{Fig.~\ref{#1}}

\newcommand\bsm{\bar{B}_s\to\mu^+\mu^-}

\newcommand\bsg{\bar{B}\to X_s\gamma}

\newcommand\dll{\delta_{LL}^{d}} \newcommand\drr{\delta_{RR}^{d}}
\newcommand\dlr{\delta_{LR}^{d}} \newcommand\drl{\delta_{RL}^{d}}

\newcommand\dxy{\delta_{XY}^{d}}

\newcommand\mgl{m_{\widetilde{g}}}
\newcommand\msq{m_{\widetilde{q}}}

\newcommand\brbsm{\mathrm{BR}(\bar{B}_s\to\mu^+\mu^-)}

\newcommand\brbsg{\mathrm{BR}(\bar{B}\to X_s\gamma)}
\newcommand\delmbs{\Delta M_{B_s}}

\newcommand\tanb{\tan\beta}

\newcommand\suthree{\mathrm{SU}(3)}

\newcommand\xdl{x_{\widetilde{d}_L}}     \newcommand\xdr{x_{\widetilde{d}_R}}
\newcommand\xdrl{x_{\widetilde{d}_{RL}}}

\newcommand\yul{y_{\widetilde{u}_L}}     \newcommand\yur{y_{\widetilde{u}_R}}

\newcommand\epst{\epsilon_3}
\newcommand\epsg{\epsilon_s}

\newcommand\epsy{\epsilon_Y}

\newcommand\BLOfact{1+\epst\tanb}
\newcommand\BLOfacg{1+\epsg\tanb}

\newcommand\ps{\,\mbox{ps}}
\newcommand\tev{\,\mbox{TeV}}
\newcommand\gev{\,\mbox{GeV}}
\newcommand\mev{\,\mbox{MeV}}

\newcommand\etc{\textit{etc.}}
\newcommand\ie{\textit{i.e.}}

\begin{document}

\begin{titlepage}
\pagestyle{empty}
\baselineskip=21pt
\rightline{KYUSHU--HET--94}  
\vskip 1.5cm
\begin{center}

{\Large\bf  New Constraints on SUSY Flavour Mixing\\ in Light of
  Recent Measurements at the Tevatron}

\end{center}   
\begin{center}   
\vskip 0.75 cm
{\bf John Foster}${}^a$\,\footnote{\, john.foster@pd.infn.it}, 
{\bf Ken-ichi Okumura}${}^b$\,\footnote{\, okumura@higgs.phys.kyushu-u.ac.jp} 
and {\bf Leszek Roszkowski${}^c$}\,\footnote{\, L.Roszkowski@sheffield.ac.uk}
\vskip 0.1in
\vskip 0.4cm ${}^a$ {\it Dipartimento di Fisica, Universit\`a di
Padova and INFN Sezione di Padova,\\ Via F. Marzolo 8, I--35131,
Padua, Italy}\\ 
${}^b$ {\it Department of Physics, Kyushu University\\
Fukuoka 812--8581, Japan}\\ 
${}^c$ {\it Department of Physics and
Astronomy, University of Sheffield\\ Sheffield S3 7RH, UK}

\vskip 1cm
\abstract{We discuss the implications, for general flavour mixing, of
  the recent results regarding the new allowed range for mixing in the
  $B_s$ meson system, as well as the new improved bounds on
  $\brbsm$. Constraints on right handed insertions, in particular,
  improve considerably in the large $\tanb$ regime where the
  contributions to $B_s$ mixing are dominated by double Higgs
  penguins. Similarly, the allowed regions of parameter space when
  varying multiple insertions also decrease significantly.}
\end{center}
\baselineskip=18pt \noindent

\vfill
\end{titlepage}

\section{Introduction}

With the high precision data currently being taken at the B factories
and the Tevatron, and the era of the LHC approaching, questions
regarding constraining the flavour structure of new physics are now
becoming especially pertinent.  One model of new physics where the
flavour structure can be extremely varied is the Minimal
Supersymmetric Standard Model (MSSM). In this model the soft terms
that are introduced by supersymmetry (SUSY) breaking can, in
principle, be arbitrary matrices in flavour space and lead to
potentially catastrophic contributions to well understood processes
such as $\bsg$.  The way this problem is often addressed is to assume
that the soft SUSY breaking terms obey the constraints imposed by
minimal flavour violation (MFV)~\cite{MFV}. In this framework, the
only source of CP and flavour violation in the MSSM is the CKM matrix
$K$. While this might initially seem to be rather appealing, it is
only an assumption, and it should be confronted with data. Indeed
verifying (or falsifying) this assumption would provide a key insight
into the possible nature of SUSY breaking.

The means by which one can probe flavour violating effects beyond the 
MFV scenario is provided by the general flavour mixing (GFM) framework.
In GFM the soft terms are treated as being basically arbitrary, with 
the various flavour violating entries constrained by the currently
available experimental data. Variations from MFV are then parameterised
by the dimensionless variables $\dxy$~\cite{GFMorig}
\begin{align}
\left(\dll\right)_{ij}=
&\frac{\left(m_{d,LL}^2\right)_{ij}}
{\sqrt{\left(m_{d,LL}^{2}\right)_{ii}\left(m^{2}_{d,LL}\right)_{jj}}},
&\left(\dlr\right)_{ij}=
&\frac{\left(m_{d,LR}^2\right)_{ij}}
{\sqrt{\left(m_{d,LL}^{2}\right)_{ii}\left(m^{2}_{d,RR}\right)_{jj}}},
\label{GA:dels1}
\\
\left(\drl\right)_{ij}=&
\frac{\left(m_{d,RL}^2\right)_{ij}}
{\sqrt{\left(m_{d,RR}^{2}\right)_{ii}\left(m^{2}_{d,LL}\right)_{jj}}},
&\left(\drr\right)_{ij}=
&\frac{\left(m_{d,RR}^2\right)_{ij}}
{\sqrt{\left(m_{d,RR}^{2}\right)_{ii}\left(m^{2}_{d,RR}\right)_{jj}}}.
\label{GA:dels2}
\end{align}
where $i,j=1,2,3$ and $m^2_{d,XY}$ ($X,Y=L,R$) are related to the
conventional soft--terms $m^2_Q$, $m^2_D$,~\etc, by unitary
transformations that rotate the quark fields from the interaction
basis to the so--called physical super CKM basis where the quark mass
terms are diagonal in flavour space. More details of the notation and
conventions we use are presented in~\cite{OR,FOR,FOR3}.

The aim of this Letter is to discuss the new bounds on the flavour violating
entries relevant to $b\to s$ transitions in light of the recent
measurements of $B_s$ mixing made by the D\O~and CDF~Collaborations at
the Tevatron~\cite{dzero-dmb,cdf-dmb}. 
Recently the D\O\ Collaboration has made a preliminary announcement of an upper bound on the
parameter $\delmbs$ with an allowed range of~\cite{dzero-dmb}
\begin{align}
17\ps^{-1} < \left(\delmbs\right)_{\mathrm{D\O}} <21\ps^{-1}
\label{BBexpd0}
\end{align}
at the 90\% confidence limit. Subsequently the result has been dramatically improved
by the CDF Collaboration  with the measurement~\cite{cdf-dmb}
\begin{align}
\left(\delmbs\right)_{\mathrm{CDF}}=17.33^{+0.42}_{-0.21}\pm 0.07\ps^{-1},
\label{BBexpcdf}
\end{align}
where the first error is statistical and the second systematic. 

Both results are in reasonable (within $2\sigma$) agreement with the
SM predictions resulting either from fits to the unitarity
triangle~\cite{SMpred} 
\begin{align}
\left(\delmbs\right)_{\mathrm{CKMfitter}}&=21.7^{+5.9}_{-4.2}\ps^{-1},&
\left(\delmbs\right)_{\mathrm{UTfit}}&=21.5\pm 2.6\ps^{-1},
\end{align}
or from a direct SM calculation~\cite{SMpred2} (for a discussion
of the current status of this prediction and the impact of
the various lattice inputs, see~\cite{SMpred3}).\footnote{Using the lattice
inputs described in~\cite{Hashimoto} we obtain a central value of $18.9\ps^{-1}$
for the SM contribution that is used in the forthcoming plots.}

With the strong evidence for an upper bound on $\delmbs$ mixing it is 
interesting to consider the impact this new constraint will have when
constraining general SUSY models. 

The supersymmetric corrections to $\delmbs$ arise at the one loop
level from box diagrams involving the exchange of charginos, gluinos,
neutralinos and squarks, as well as contributions arising from charged
Higgs exchange~\cite{dmbsusy}. The supersymmetric contributions, which
at one loop are the only ones that are sensitive to the flavour
structure of SUSY breaking, are typically dominated by the
$\alpha_s$--enhanced gluino diagrams in the GFM framework. The
corrections arising from charginos and neutralinos can therefore be
neglected in regions of parameter space where $\tanb$ is
small. However, as $\tanb$ increases, the two contributions can become
increasingly important and must be taken into account in a consistent
analysis of the bounds on the insertions.  In addition to these
diagrams, at large $\tanb$ a new two loop contribution arises due to
double Higgs penguins mediated by neutral
Higgs~\cite{DHP,BSM,BCRS,BLO,FOR}. These contributions benefit from an
enhancement by $\tan^4\beta$ and vary only as $1/m_A^2$ (the mass of
the pseudoscalar Higgs $A^0$). When compared to the $1/\msq^2$
dependence (where $\msq$ is the squark mass scale) of the SUSY
mediated box diagrams it is apparent that, when $\tan\beta$ is large,
the double Higgs penguin contributions can play a significant r\^ole
and even dominate the dynamics associated with the new physics
contributions to $B_s$ mixing.

Of course, $B_s$ mixing is not the only possible constraint on the
supersymmetric contributions to $b\to s$ transitions. As discussed
in~\cite{FOR,FOR3} both $\bsg$ and $\bsm$ can also play a r\^ole in
this endeavour.  The decay $\bsg$, for instance, has remained a
mainstay in analyses of $b\to s$ transitions for some time now and
provides quite stringent bounds on certain sources of flavour
violation in the MSSM~\cite{BSG,OR,FOR}. The decay $\bsm$, on the
other hand, currently remains unobserved, however it is intimately
linked to $B_s$ mixing as both processes can benefit from large
enhancement through neutral Higgs penguin dynamics in the large
$\tanb$ regime. The neutral Higgs penguin contribution to $\bsm$
benefits from $\tan^6\beta$ enhancement~\cite{DHP,BSM,BCRS,BLO,FOR}
and, due to the helicity suppression of the SM contribution it is
possible to induce exceptionally large contributions to the decay that
can quite easily approach the preliminary limit from the CDF
Collaboration~\cite{cdf-bsm} of
\begin{align}
\brbsm_{\mathrm{CDF}}<8.0\times 10^{-8}
\label{BSMexp}
\end{align}
at the $90\%$ confidence level. The value of the bound increases to
$1.0\times 10^{-7}$ at the $95\%$ confidence level. For reference the
SM prediction for the branching ratio is~\cite{bsm-sm}
\begin{align}
\brbsm_{\mathrm{SM}}=\left(3.46\pm 1.5\right)\times 10^{-9}.
\label{BSMsm}
\end{align}
As both $B_s$ mixing and $\bsm$ share a similar dependence
on the same underlying vertex, it is natural to ask how the new
bound on $B_s$ mixing might effect the current prospects for the detection
of $\bsm$.

In this Letter we shall perform a complete analysis
of the constraints that emerge from the new CDF bound on $\delmbs$~\eqref{BBexpcdf}.
To ensure the validity of our analysis in as wide a range of 
parameter space as possible we shall compute the contributions
arising from all the possible SUSY contributions to $B_s$ mixing that
arise at the one loop level as well as the double Higgs penguin diagram
that becomes significant at large $\tanb$.

In addition, we apply the procedure discussed in detail in~\cite{FOR}
to include the large effects that appear beyond the leading order
(BLO). These effects include terms that are enhanced by either large
logs or $\tanb$~\cite{bsgblo,BCRS,BLO,OR,FOR}. Large logs are
induced by running from the SUSY scale $\mu_{SUSY}$ to the
electroweak scale $\mu_W$.  The terms enhanced by $\tanb$, on the
other hand, arise from threshold corrections to the down quark mass
matrix~\cite{QM,FOR} and Higgs
vertices~\cite{CH,bsgblo,BSM,DHP,BCRS,BLO,FOR}.  The $\tanb$ enhanced
terms can be especially important, and their inclusion can lead to
large differences between a purely LO calculation and one that
properly takes into account BLO effects~\cite{OR,FOR}. In particular,
the inclusion of such BLO corrections leads to a focusing effect
that, at large $\tanb$ and $\mu>0$, can significantly loosen the
bounds on SUSY sources of flavour violation~\cite{OR,FOR}.

Other recent analyses have also discussed the impact of the recent
measurement~\eqref{BBexpcdf} both in the context of MFV~\cite{dmbrecentMFV} and
GFM~\cite{dmbrecentGFM,SMpred3}. We should point out, however, that, in
contrast to the analyses presented in~\cite{dmbrecentGFM,SMpred3}, we
take into account all the relevant SUSY one--loop diagrams as well as the
double Higgs penguin diagram that appears at the two--loop level. Possible enhancements
to the phase associated with $B_s$ mixing have also been discussed within the context of this new
constraint~\cite{dmbphase}. In this analysis, however, we shall treat
the various sources of SUSY flavour violation as real and any
contributions to the phase of $B_s$ mixing would therefore be expected
to be small in this limit.

The constraints we impose take the following form. For $\bsg$ we require
that the prediction for $\brbsg$ lies between the bounds
\begin{align}
2.65\times 10^{-4}<\brbsg<4.35\times 10^{-4},
\end{align}
a range that can be obtained by adding the SM and experimental errors in quadrature and taking the $2\sigma$ interval around the
most recent experimental world average of $(3.55\pm 0.24^{+0.09}_{-0.10}\pm 0.03)\times 10^{-4}$~\cite{HFAG} (compared to $(3.57\pm 0.30)\times 10^{-4}$ in the SM~\cite{SM:bsg}). For $\bsm$ we require
that $\brbsm$ satisfies the $90\%$ confidence interval~\eqref{BSMexp}
and, finally, for $\delmbs$ we impose the constraint
\begin{align}
12.53\ps^{-1}<\delmbs<22.13\ps^{-1}
\end{align}
with the allowed range obtained by taking the $1\sigma$ interval after combining the experimental and SM errors in quadrature. This requirement, in particular
is especially strict, however, it provides an appreciation of the
possible implications of the new results concerning $B_s$ mixing.

Finally, while we parameterise the amount of flavour violation
using $\dxy$ we should emphasise that all of the forthcoming
calculations are performed in the mass eigenstate basis, rather
than the  mass insertion approximation that can become inaccurate
for large values of $\dxy$.

\section{Constraints on Single Insertions}

Let us first consider the constraints that can be placed on single insertions.
As discussed in the previous section, the contributions to $\delmbs$
in the large $\tanb$ regime are dominated by double Higgs penguins
that benefit from an enhancement by $\tan^4\beta$, as well as a reduced
mass suppression, compared to the gluino mediated box diagrams that are
the dominant contribution at low $\tanb$. Using the results
concerning the neutral Higgs penguin gathered in~\cite{DHP,BCRS,BLO,FOR}
it is relatively easy to evaluate the contributions to $\delmbs$ arising
from these diagrams
\begin{align}
\delmbs^{DP}=&-\frac{8 G_F}{3 \sqrt{2}} m_{B_s} f_{B_s}^2
P_2^{LR}\frac{m^2_b\tan^4\beta}{m_A^2} 
\nonumber\\
&\times\left\{\frac{\left(\epsilon_Y Y_t^2 K_{ts}
  K^{\ast}_{tb}+\epsilon_{LL}\xdl\dll\right)}
    {\left(\BLOfact\right)\left(\BLOfacg\right)} +
    \frac{\mgl}{m_b}\frac{\epsilon_{RL}\epsilon_s\xdrl\dlr}{\left(\BLOfacg\right)} + 
    \frac{m_s}{m_b}\frac{\epsilon_{RR}\xdr\drr} {\left(\BLOfacg\right)^2}\right\}
\nonumber\\
&\times\left\{
\frac{m_s}{m_b}\frac{\left(\epsilon_Y Y_t^2 K_{ts} K^{\ast}_{tb} +
  \epsilon_{LL}\xdl\dll\right)}{\left(\BLOfact\right)\left(\BLOfacg\right)} 
+\frac{\mgl}{m_b}\frac{\epsilon_{RL}\epsilon_3\xdrl\drl}{\left(\BLOfact\right)}
+\frac{\epsilon_{RR}\xdr\drr}{\left(\BLOfact\right)^2}
\right\},
\label{EQ1}
\end{align}
where $f_{B_s}$ and $P_2^{LR}$ (see~\cite{dmbmast} for a definition) are determined from the lattice and are
the principle source of error associated with the calculation for
$\delmbs$.  Aside from $\tanb$ and $m_A$ the various SUSY parameters
only enter as ratios encoded in the factors of $\epsilon$ that are
gathered in the appendix. This fact gives rise to a non--decoupling
effect where the double penguin contribution to $\delmbs$ can be
sizable, even for multi--$\tev$ scale sparticle masses, provided that
the Higgs sector remains relatively light.  The BLO corrections to the
above expression take the form of the factors of $\BLOfact$ and
$\BLOfacg$ that appear in the denominators of each term. For $\mu>0$
($\mu<0$) these factors typically act to suppress (enhance) the double
Higgs penguin contribution. In addition to these corrections the
insertions $\dlr$ and $\drl$ also appear in the above expression once
BLO effects are taken into account~\cite{FOR}. 

Numerically, for
degenerate sparticle parameters of $1\tev$ the contribution to $\delmbs$
becomes
\begin{align}
\delmbs^{DP}=&-18\ps^{-1}\left(\frac{f_{B_s}}{230\mev}\right)^2
\left(\frac{P_2^{LR}}{2.56}\right)\left(\frac{\tanb}{40}\right)^4 
\left(\frac{500\gev}{m_A}\right)^2 
\nonumber\\
&\times\left\{
\frac{\left(1-33.6\,\dll\right)}{\left(1\pm
  0.38\frac{\tanb}{40}\right) \left(1\pm 0.29\frac{\tanb}{40}\right)}
+\frac{385.2\,\dlr}{\left(1\pm 0.38\frac{\tanb}{40}\right)}
-\frac{0.75\,\drr}{\left(1\pm 0.29\frac{\tanb}{40}\right)^2}
\right\}
\nonumber\\
&\times\left\{
\frac{0.022\left(1-33.6\,\dll\right)}{\left(1\pm
  0.38\frac{\tanb}{40}\right) \left(1\pm 0.29\frac{\tanb}{40}\right)}
+\frac{290.6\,\drl}{\left(1\pm 0.29\frac{\tanb}{40}\right)}
-\frac{33.6\,\drr}{\left(1\pm 0.29\frac{\tanb}{40}\right)^2}
\right\},
\label{EQ2}
\end{align}
where we have assumed that the trilinear soft mass term $A_u$ has the
opposite sign compared to $\mu$ to ensure compatibility with the constraint
supplied by $\bsg$.  The sign of the contributions that
appear in the denominator of each term is determined by
$\mathrm{sgn}(\mu)$.  It should be noted that the seemingly large
contributions arising from LR and RL insertions to each Higgs penguin
are tempered by the fact that these insertions scale as
$1/M_{\mathrm{SUSY}}$ and are, therefore, at least one or two orders
of magnitude smaller than LL and RR insertions.

Inspection of~\eqs{EQ1}{EQ2} reveals that in the limit of MFV, or when
only an LL or LR insertion is non--zero, the double Higgs penguin
contribution to $\delmbs$ is suppressed by $m_s/m_b$. In addition, the
contributions to $\delmbs$ arising from MFV effects and LL insertions
destructively interfere with the SM value driving it towards the
current lower bound on the parameter~\cite{DHP,BCRS,BLO}. As such, the
impact of the new upper bound in constraining both MFV and LL or LR
insertions tends to be negligible once one takes into account the
constraints arising from $\bsg$ and $\bsm$.

It is evident from the above formulae, however, that GFM effects allow
one to avoid the suppression by $m_s/m_b$ in the presence of either RL
or RR insertions. If only these insertions are non--zero (see the next
section for a discussion regarding the more general case) the
suppression by the strange quark mass is overcome via the interplay
between the MFV contribution to one of the neutral Higgs penguins and
the contribution arising from either one of these insertions.  This
effect, which was first pointed out in~\cite{FOR}, can give to rise to
extremely large contributions to $\delmbs$ while still being in
agreement with the current bounds on $\bsm$ and $\bsg$. In addition,
the sign of the correction to $\delmbs$ depends on the sign of
the RL or RR insertion. Consequently, it is possible to generate
contributions to $B_s$ mixing that interfere constructively with the
SM result and any upper bound on $B_s$ mixing will, therefore, provide
a useful constraint on these two insertions in the large $\tanb$
regime. One should note that such a contribution would be absent in a
simplistic analysis in which only GFM effects arising from gluinos
were taken into account. It is therefore vital to properly address the
contributions arising from electroweak (EW), and therefore MFV,
effects arising from higgsino exchange in any analysis placing bounds
on these two insertions in the large $\tanb$ regime.

Finally, let us point out that the formulae in~\eqs{EQ1}{EQ2} are only
approximate and do not include, for instance, the additional EW
corrections arising from gaugino exchange, which can, in some
instances, have up to a 20\% effect~\cite{BCRS,FOR}. All such corrections
are taken into account in our numerical analysis.

\begin{figure}[!tb]
\begin{center}
\begin{tabular}{c c}
	\includegraphics[width=0.45\textwidth]{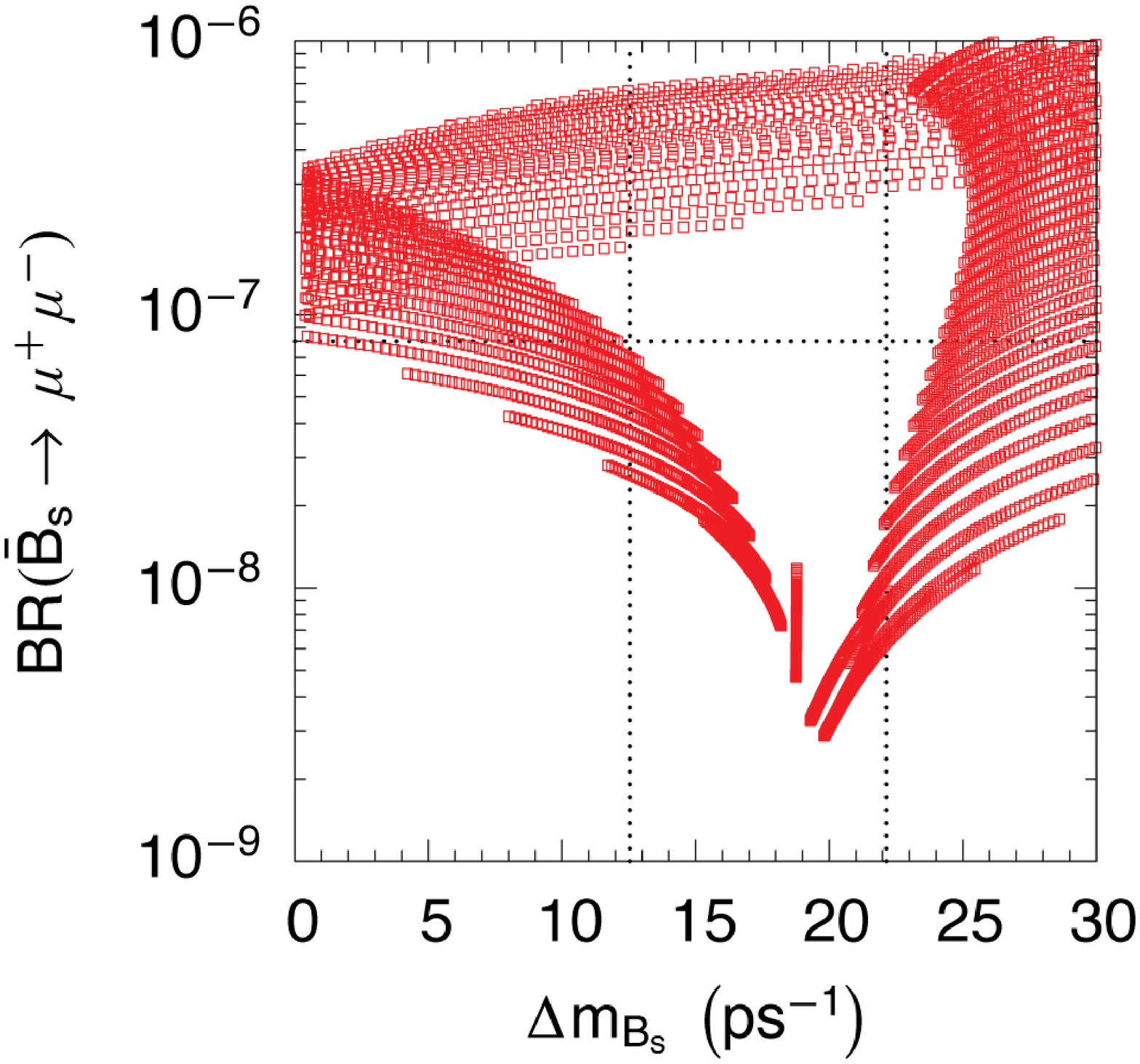}
&	\includegraphics[width=0.45\textwidth]{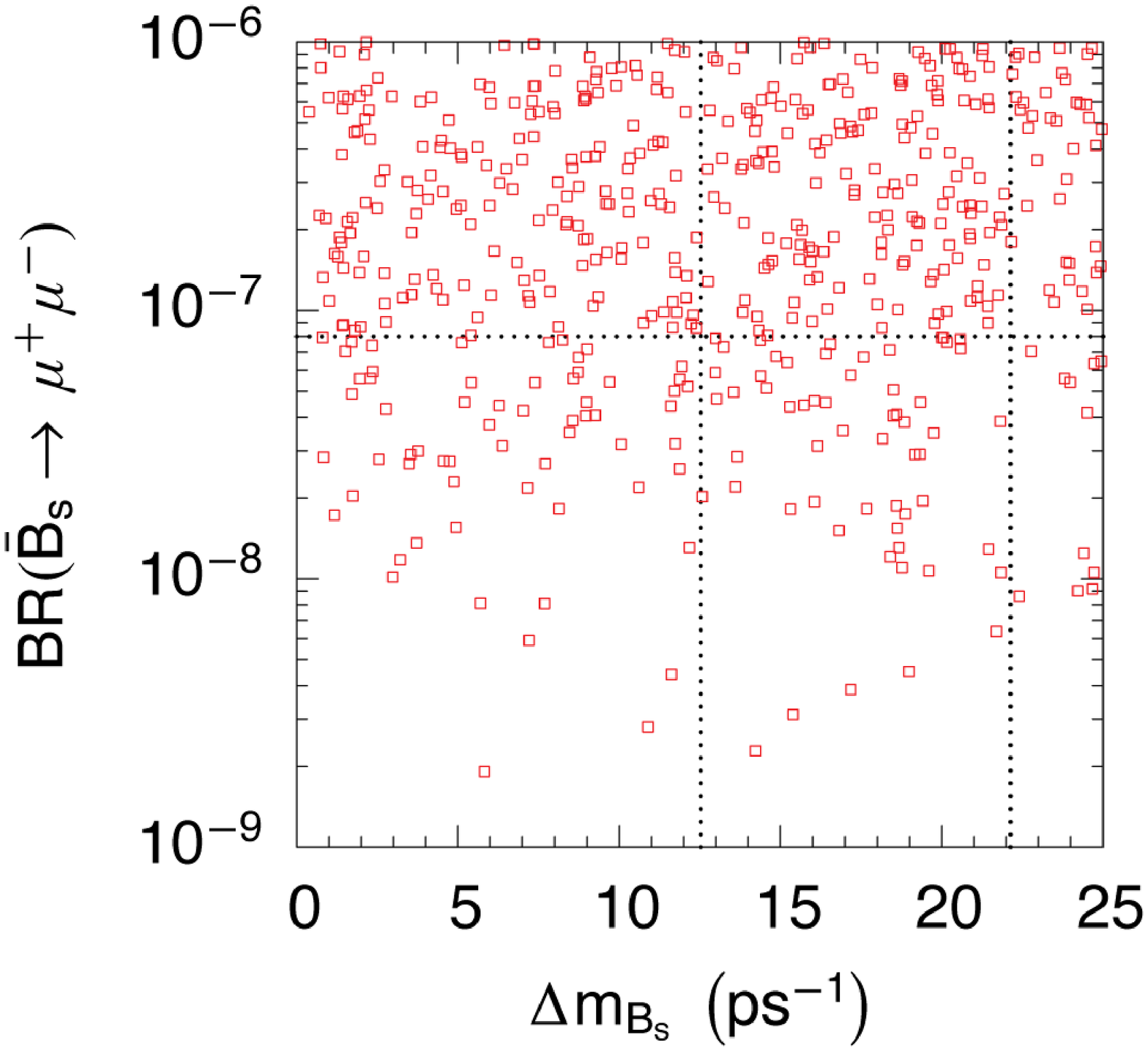}
\end{tabular}
\end{center}
\caption{ Scatter plots illustrating the correlation between $\brbsm$
and $\delmbs$.  In the panel to the left the insertion $\drr$ is
scanned over the interval $[-0.8,0.8]$ and $\msq$ is varied over the
range $[500,1500]\gev$ the remaining parameters are given by
$\mu=m_A=-A_u=500\gev$, $\mgl=1\tev$. In the panel on the right all
four insertions are varied, with $\dll$ and $\drr$ over the range
$[-0.8,0.8]$, and $\dlr$ and $\drl$ over the range $[-0.08,0.08]$.
The soft sector is described by $\msq=-A_u=\mgl/\sqrt{2}=\sqrt{2}\mu=2
m_A=1\tev$. In both panels $\tanb=40$. Points inconsistent with the
constraint supplied by $\bsg$ are not shown.
\label{FIG0}
}
\end{figure}
Before discussing the parameter dependence of the new bounds on $B_s$
mixing and $\bsm$ let us briefly comment on the correlation between
the two observables.
Such a situation is illustrated in~\fig{FIG0}
where, in the panel on the left, we show the correlation between
the $\brbsm$ and $\delmbs$ for varying $\drr$ (one can find a similar
plot for $\dll$ insertions in Fig. 26 of~\cite{FOR}). As is evident
from the plot, contributions to $\bsm$ arising from these insertions
are now severely constrained by the new bounds on $B_s$ mixing.
It is well known that varying all four insertions at the same time
essentially destroys a large amount of the correlation between the 
two variables and this is illustrated in the figure to the right.
However, it is also apparent that both bounds on $B_s$ mixing
and $\brbsm$ rule out a large proportion of the available
parameter space in this scenario. We shall discuss the limits on 
multiple insertions in the next section.

\begin{figure}[!p]
\begin{center}
\begin{tabular}{c c}
	\includegraphics[width=0.45\textwidth]{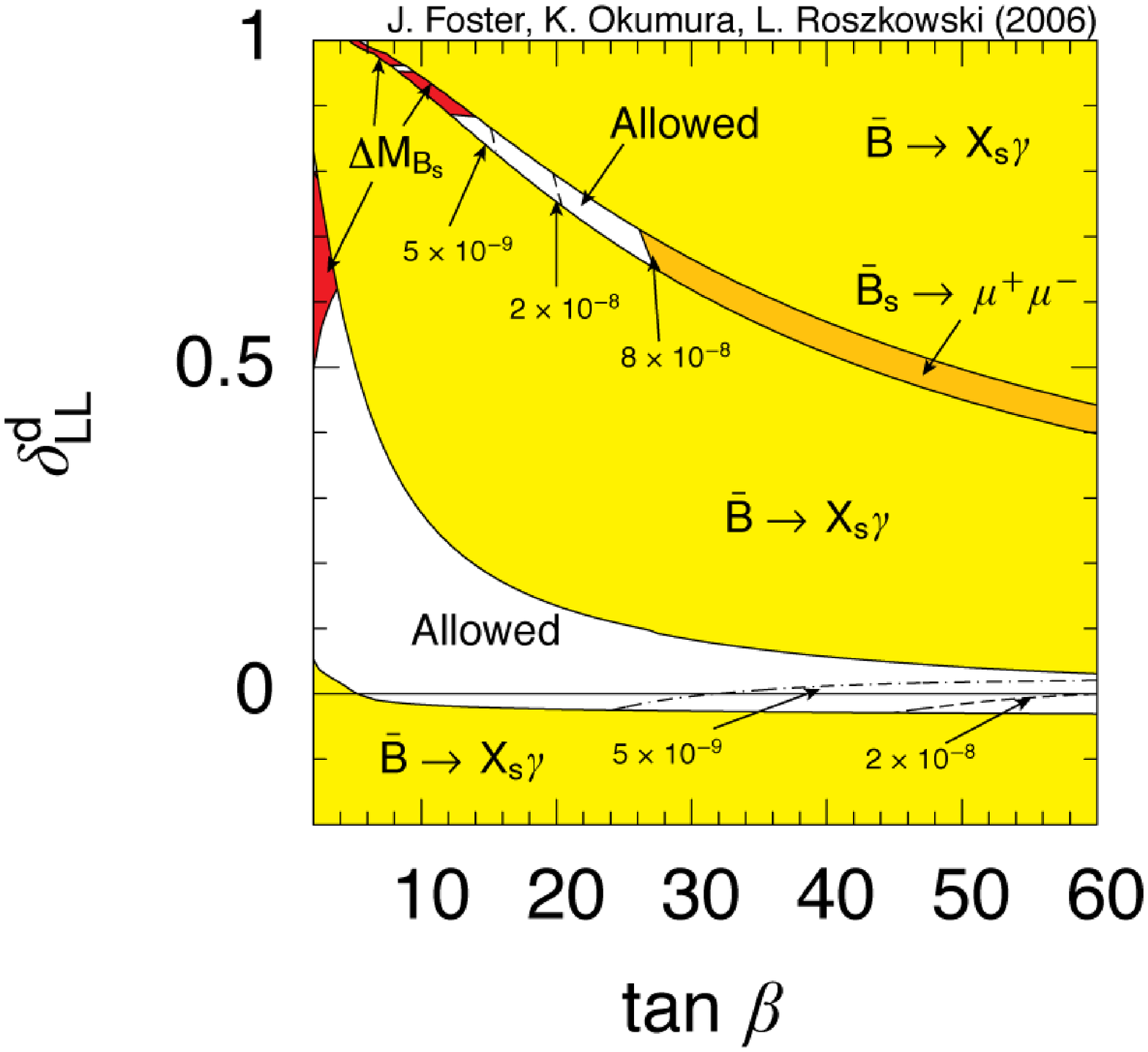}
&	\includegraphics[width=0.465\textwidth]{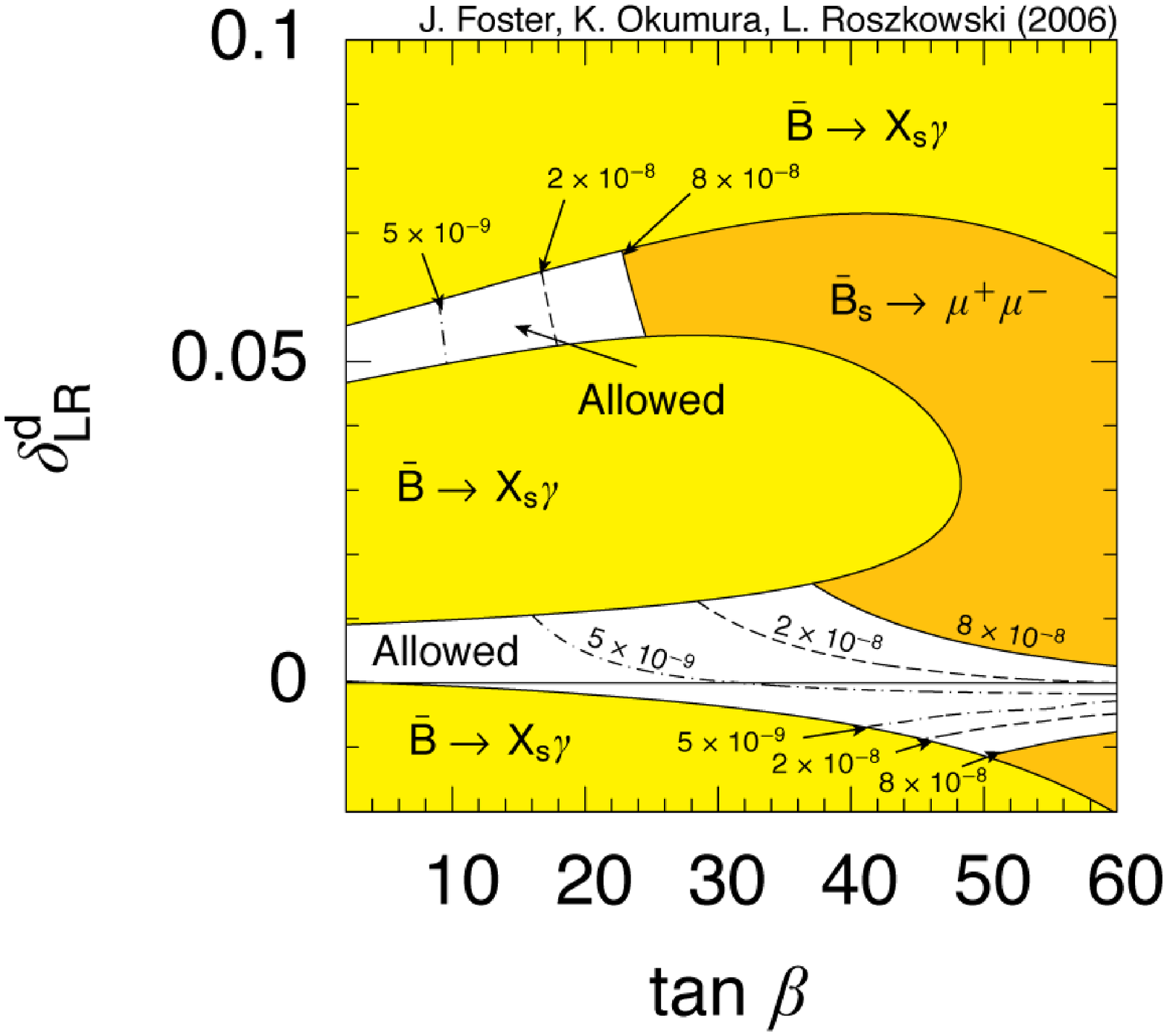}\\
	\includegraphics[width=0.45\textwidth]{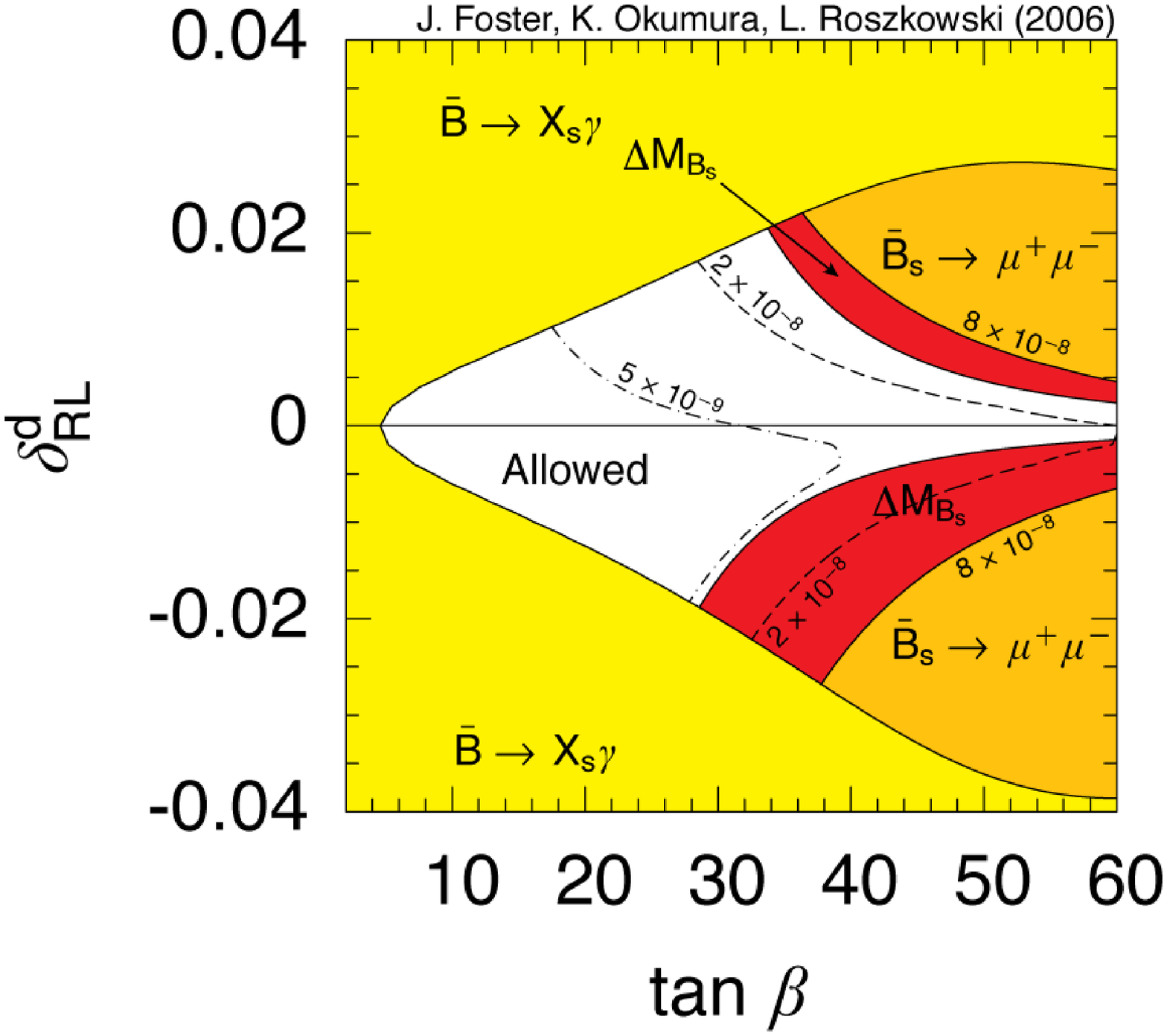}
&	\includegraphics[width=0.44\textwidth]{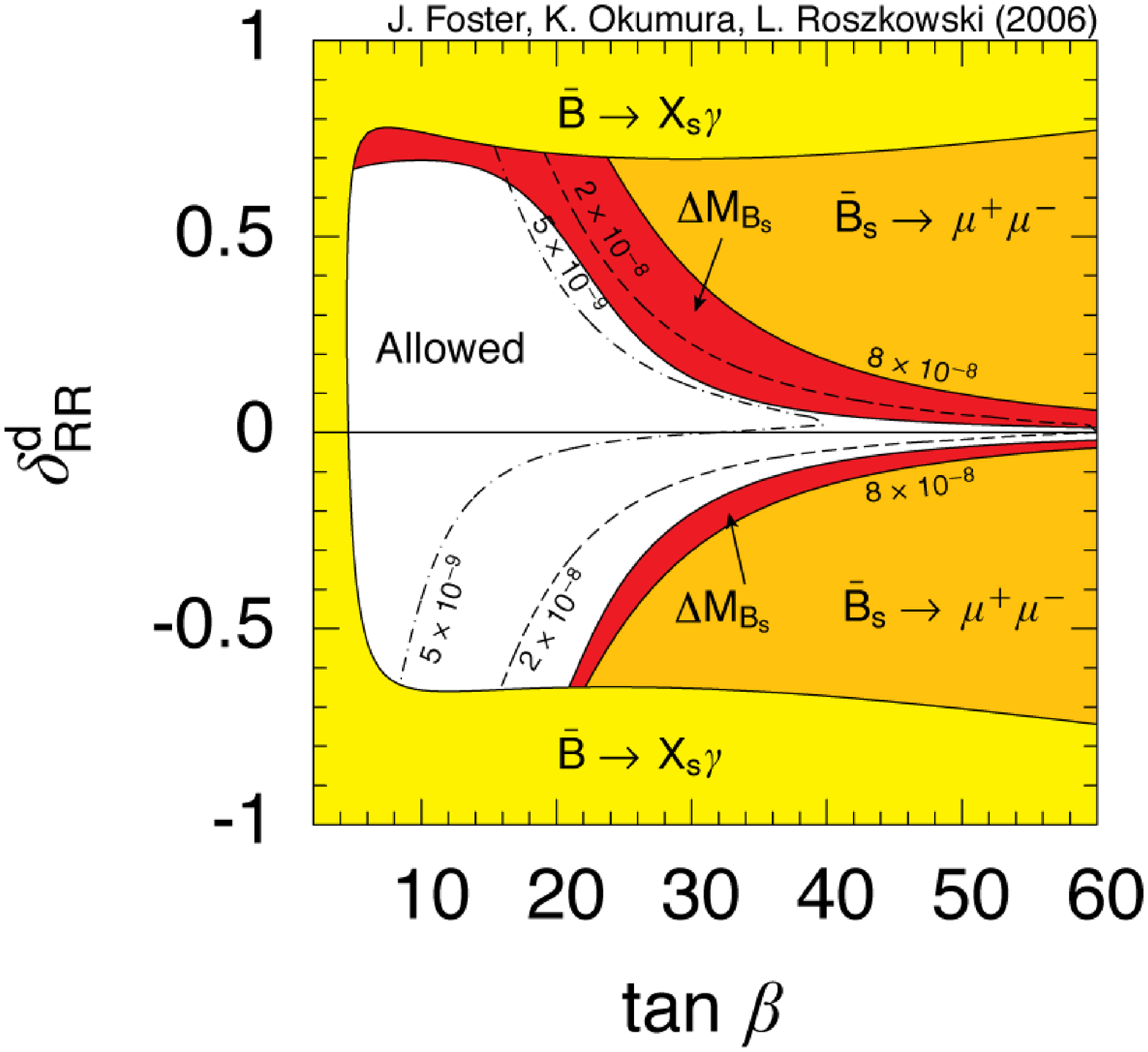}
\end{tabular}
\end{center}
\caption{Contour plots showing the limits on the GFM parameters $\dxy$
  for varying $\tanb$. In each panel only one $\dxy$ is varied and
  the rest are put to zero. The soft sector is parameterised as
  follows $\msq=\mgl=1\tev$ and $\mu=-A_u=500\gev$ and
  $m_A=500\gev$. Regions excluded by $\bsg$ (\ie~outside of the region
  $2.65\times 10^{-4} < \brbsg < 4.35\times 10^{-4}$) are shaded in yellow
  (light grey). The subsequent regions that are excluded by the CDF
  limit of $8\times 10^{-8}$ on $\bsm$ are shaded in orange (medium
  grey). The remaining regions that are excluded by the D\O\ and
  CDF~results on $\delmbs$ are shaded in red (dark grey). It should be
  noted that we relax the allowed region to
  $12.53\ps^{-1}<\delmbs<22.13\ps^{-1}$ to take into account the errors
  associated with the lattice inputs required to evaluate $\delmbs$.
  Finally, contours depicting values of $2\times 10^{-8}$ and $5\times
  10^{-9}$ for $\brbsm$ are shown and are delineated by dashed and
  dot--dashed lines respectively.\label{FIG1} }
\end{figure}
\fig{FIG1} shows the $\tanb$ dependence of the new constraints arising
from the new upper bound on $B_s$ mixing as well as the improved bound
on $\bsm$~\eqref{BSMexp}. The top two panels of the figure illustrate
the bounds imposed on $\dll$ and $\dlr$ insertions. In these two cases
the effect of the new bound on $B_s$ mixing is relatively minor.  The
constraint on $\dll$ insertions, for instance, only eliminates an
extreme region of parameter space at low $\tanb$ where the amplitude
for $\bsg$ has effectively flipped sign, while for LR insertions that
constraint has no effect on the bounds whatsoever. From these two
panels it is also possible to appreciate the effect that improving the
bound on $\bsm$ might have when constraining the insertions. For
instance in both panels it is possible to see that improving the bound
to $2\times 10^{-8}$ (a conservative value that is achievable at the
Tevatron) would start playing an important role in constraining the
insertions at large $\tanb$.  A bound of $5\times 10^{-9}$ (thereby
ruling out any large new physics contributions to $\bsm$) would
provide the bounds on the insertions for values of $\tanb$ as low as
10.

As discussed in~\cite{FOR,FOR3}, the bounds on the insertions $\drl$
and $\drr$ are far more dependent on the $\delmbs$ constraint, and
this is evident in the two lower panels. For both insertions the
bounds at large $\tanb$ are now imposed entirely by the $B_s$ mixing
constraint. In fact, for RR insertions, the bounds attributable to the
$\delmbs$ constraint are present for values of $\tanb$ as low as $10$
(for $\drr>0$) while values of
$\mathcal{O}(1)$ for the insertion are now heavily 
disfavoured in the large $\tanb$ regime.  It is also evident from the
plots that improving the limit on $\bsm$ will have relatively little
effect on the possible bounds that can be placed on the two insertions
until values approaching the SM prediction are reached. The flip side
of this argument is, of course, that if the Tevatron were to measure
$\bsm$ in the region of $3-10\times10^{-8}$ one could almost
automatically rule out large contributions from RR or RL insertions
being responsible for such an enhancement (ignoring possible fine
tuned scenarios).

\begin{figure}[!tb]
\begin{center}
\begin{tabular}{c c}
	\includegraphics[width=0.45\textwidth]{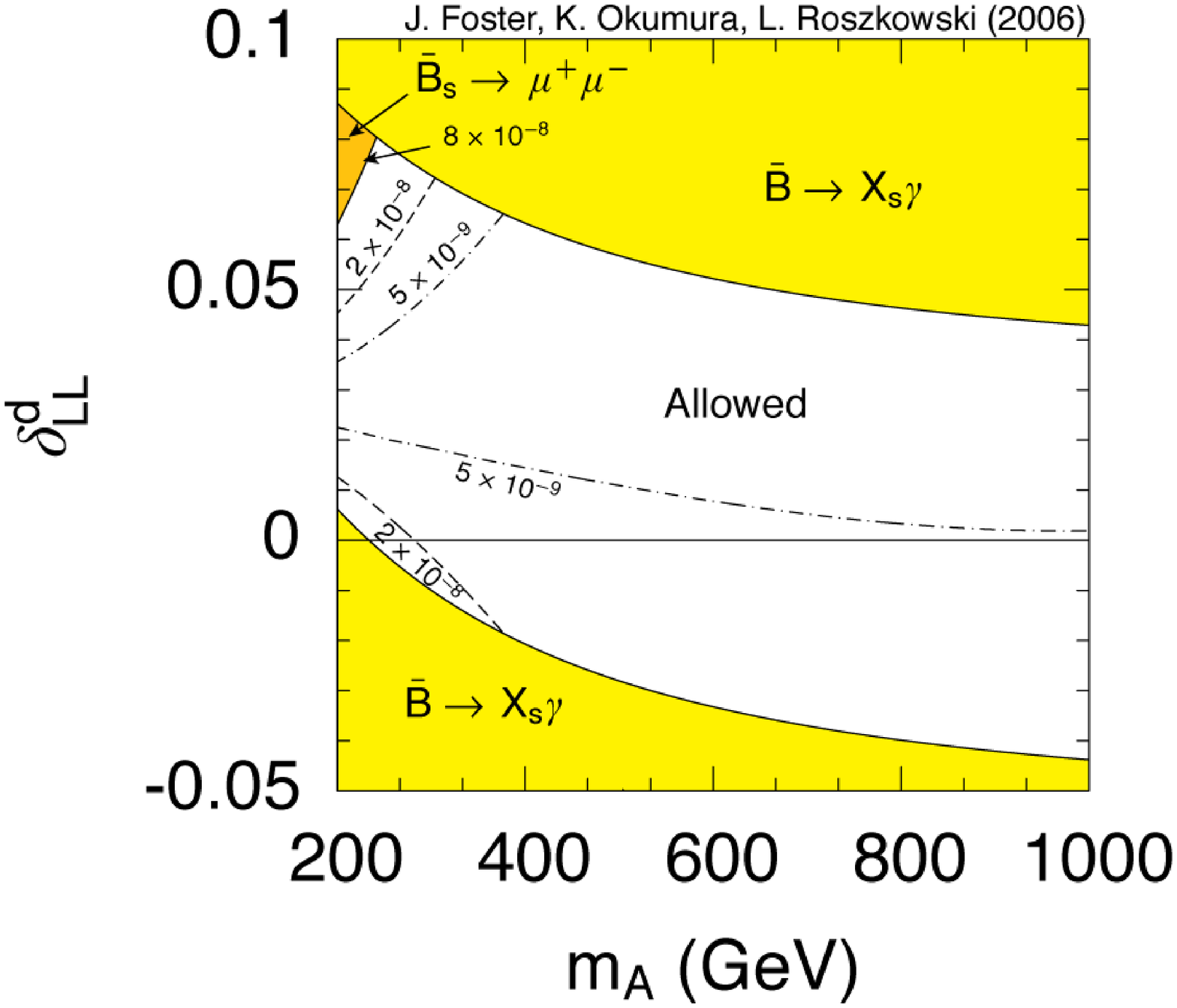}
&	\includegraphics[width=0.45\textwidth]{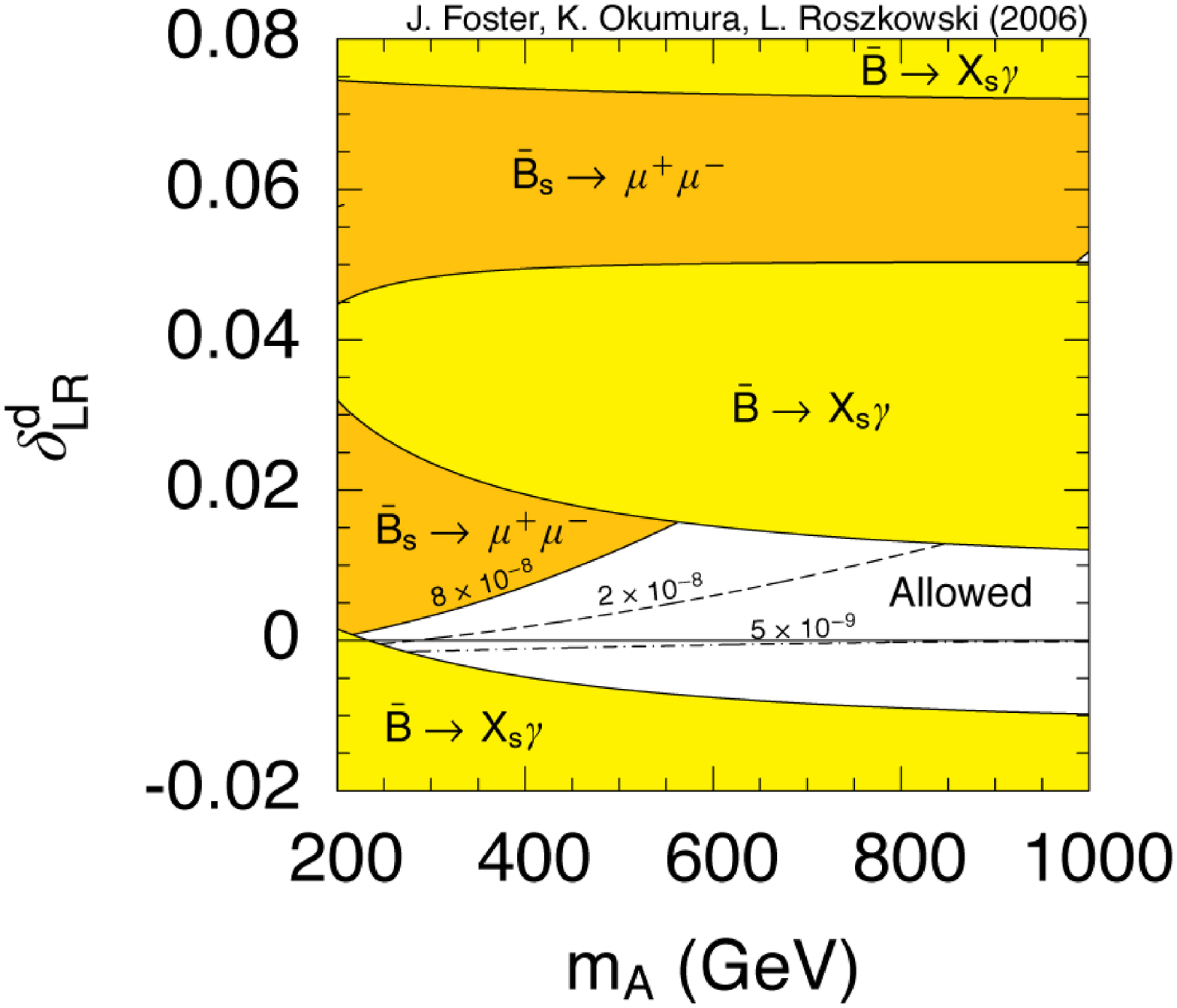}\\
	\includegraphics[width=0.45\textwidth]{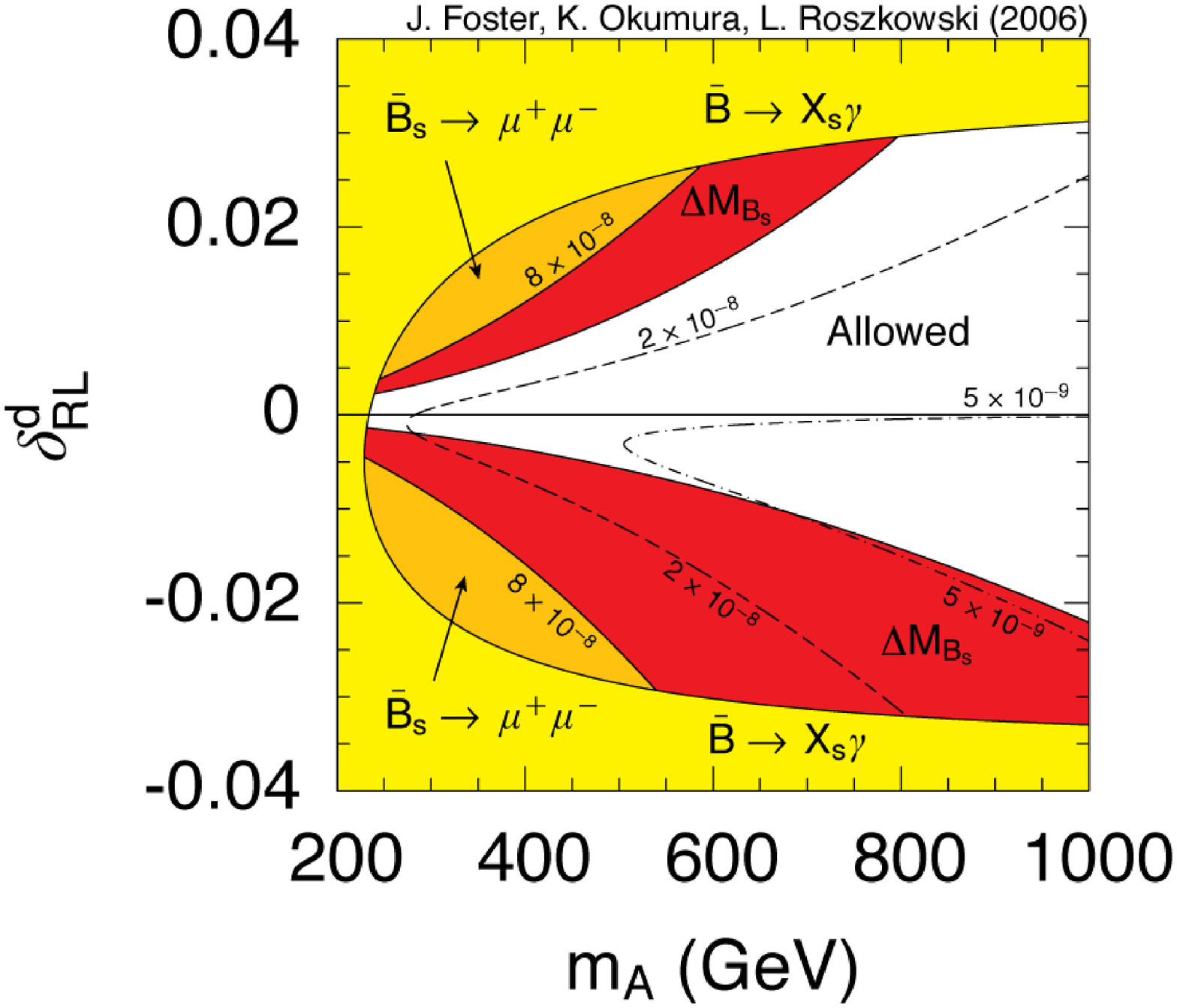}
&	\includegraphics[width=0.44\textwidth]{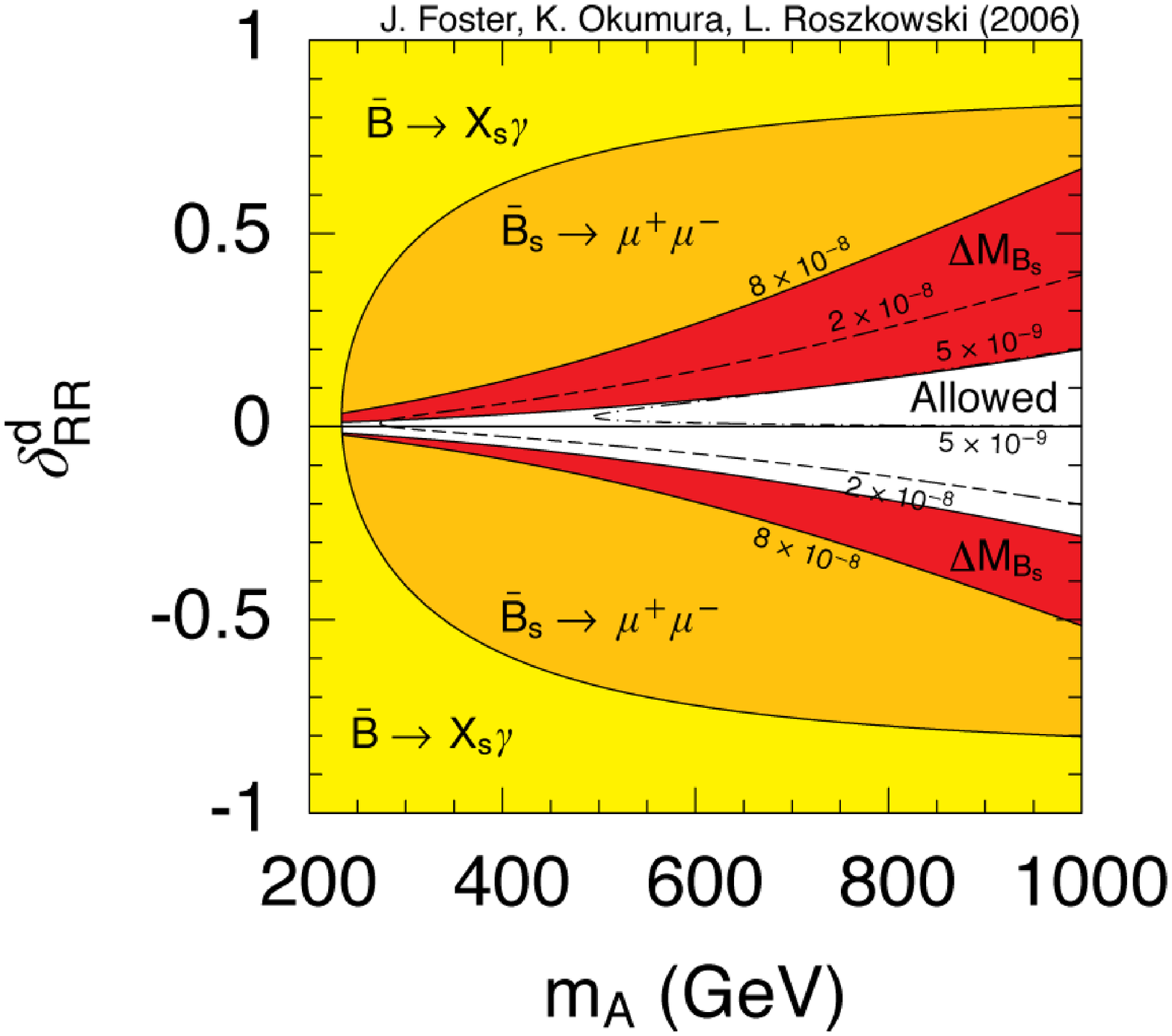}
\end{tabular}
\end{center}
\caption{
Contour plots showing the limits on the GFM parameters
  $\dxy$ for varying $m_A$. The soft sector is
  parameterised as follows $\msq=\mgl=1\tev$ and $\mu=-A_u=500\gev$ and $\tanb=40\gev$. The
  excluded and allowed regions are shaded in a similar manner to~\fig{FIG1}.\label{FIG2}
}
\end{figure}
As the constraint supplied by $B_s$ mixing also depends on $m_A$, it
is natural to ask how the bounds on the insertions change when this
parameter is varied. Such a scenario is illustrated in~\fig{FIG2}.  In
the top two panels illustrating the bounds on LL and LR insertions we
see, once again, that the new bound on $B_s$ mixing has no effect on
the bounds that can be placed on the two insertions when compared to
the existing constraints supplied by the $\Delta F=1$ processes $\bsg$
and $\bsm$. On the other hand, the impact of improving the bound on
$\bsm$ is more substantial. It is clear that any prospective
improvements will affect the low $m_A$ region of parameter space and
improve the possible bounds on the insertions substantially.

The lower two panels illustrate the bounds on the insertions $\drl$
and $\drr$.  Once again, we can see the dramatic effect the constraint
supplied by $B_s$ mixing has. Even when $m_A$ is of a similar
order as the squark masses (\ie~$1\tev$) it is clear that useful
bounds can be placed on both insertions, that surpass those derived
from either $\bsm$ or $\bsg$. In a similar manner to the plots that
appeared in~\fig{FIG1} the effect of improving the bound on $\bsm$ to
$2-3\times 10^{-8}$ on the constraints on the two insertions will be
relatively minor, however improving the limit to $5\times 10^{-9}$
would prove to be most useful in further constraining the allowed
flavour violation in the RL and RR sectors.

It is natural to consider how varying other parameters might affect
the bounds one can place on these insertions. As is evident
from~\eqref{EQ1} the remaining SUSY parameters enter as ratios encoded
in the factors of $\epsilon_i$. Therefore universally scaling the SUSY
spectrum will have relatively little effect on the bounds, provided
that $m_A$ and $\tanb$ remain constant. The dependence on single
parameters such as $\mu$ and $A_u$ can, however, be more complex.

Varying the $\mu$ parameter, for instance, generally has the effect
of increasing or decreasing the magnitude of the factors of $\epsilon$
that appear in~\eqref{EQ1}. Increasing $\mu$ initially strengthens
the bounds one can place on the insertions using $\bsm$ or $B_s$ mixing, however, once it reaches a similar order of magnitude as $\mgl$
or $\msq$ the bounds tend to remain relatively independent of further
variations in the parameter.

\begin{figure}[!tb]
\begin{center}
\begin{tabular}{c c}
	\includegraphics[width=0.45\textwidth]{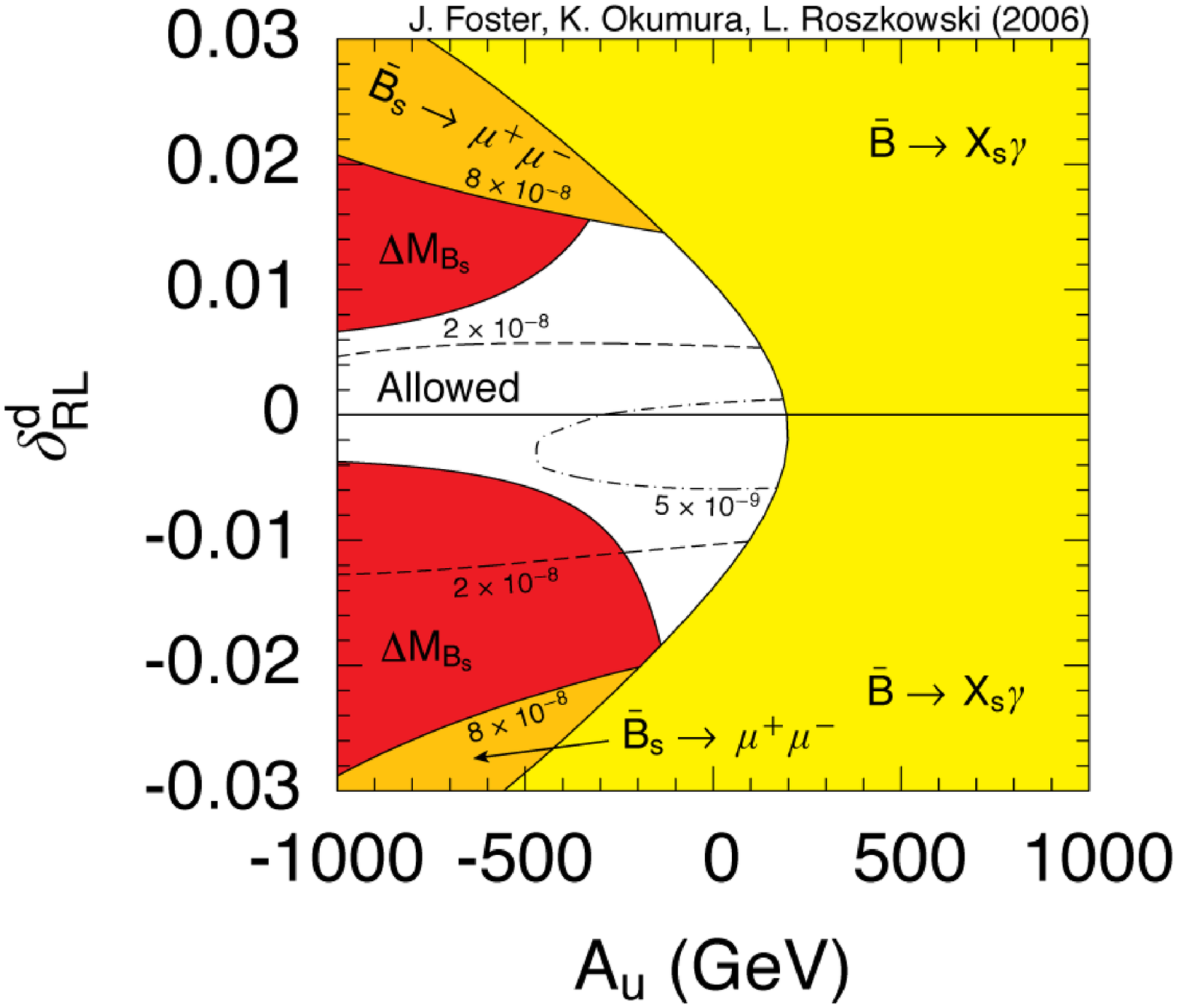}
&	\includegraphics[width=0.435\textwidth]{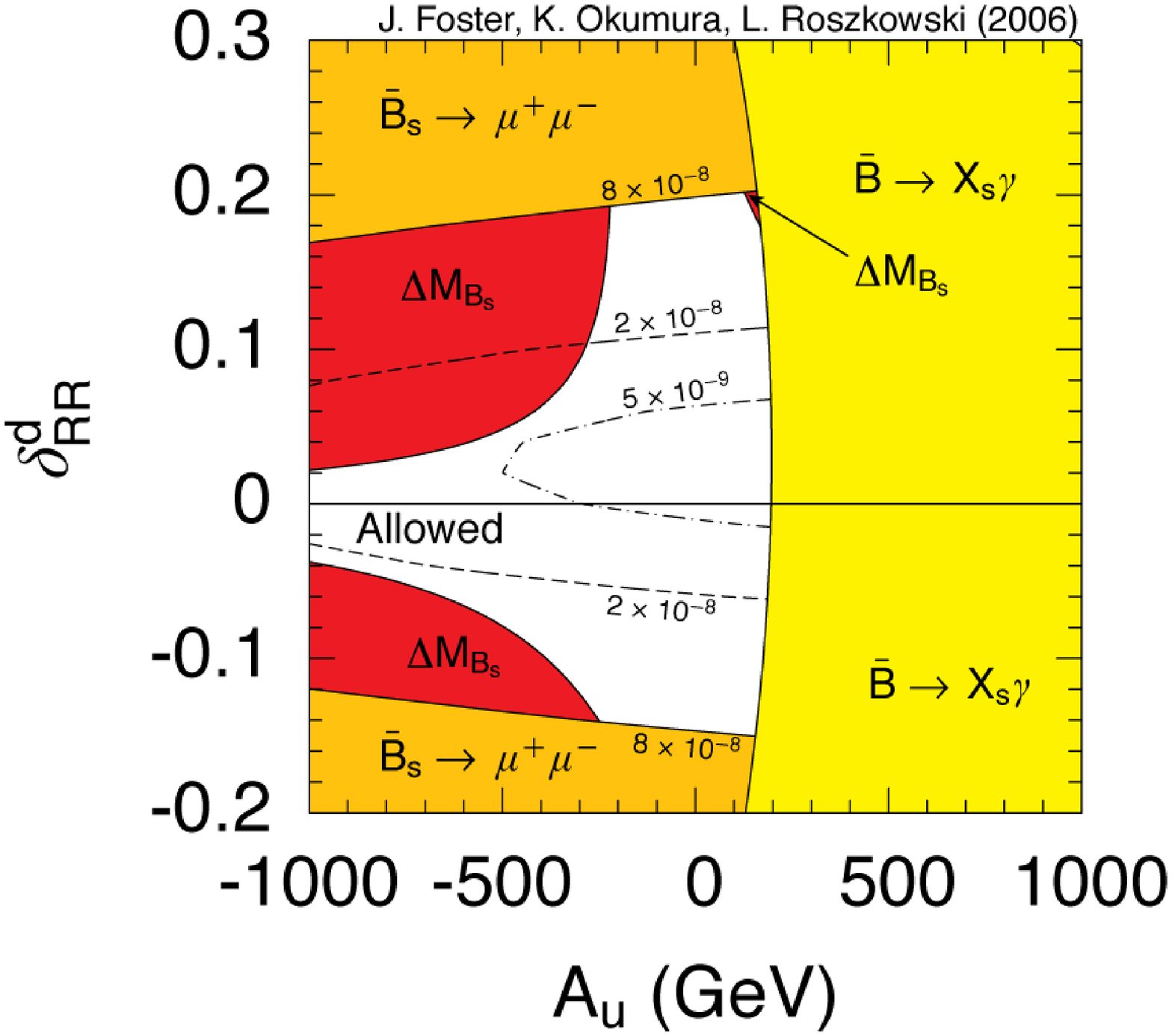}
\end{tabular}
\end{center}
\caption{ Contour plots showing the allowed ranges for $\drl$ and
$\drr$ for varying $A_u$. The soft sector is parameterised as follows
$\msq=\mgl=1\tev$ and $\mu=m_A=500\gev$ and $\tanb=40\gev$. The
excluded and allowed regions are shaded in a similar manner
to~\fig{FIG1}.\label{FIG3} }
\end{figure}
The effect of varying the parameter $A_u$ is illustrated
in~\fig{FIG3}.  $A_u$ enters into the higgsino contribution to the
neutral Higgs penguin and is encoded in the factor of $\epsilon_Y$
that appears in~\eqref{EQ1}. As such its effect on the bounds imposed
on LL or LR insertions tends to be rather small.  For RL and RR
insertions, on the other hand, the effect of varying the parameter can
be appreciably larger, as $\epsilon_Y$ (and therefore $A_u$) enters
into the double Higgs penguin contribution arising from these
insertions to $\delmbs$. Both panels in~\fig{FIG3} illustrate that
increasing the magnitude of $A_u$ tends to increase the effectiveness
of the $B_s$ mixing bound when constraining $\drl$ and $\drr$
insertions. Indeed it can be seen from both plots that, when $A_u$ is
small, the constraints on the insertions are typically replaced by
those arising from the decay $\bsm$. It should also be noted that the
bounds in each figure are relatively independent of the sign of $A_u$,
except those arising from $\bsg$ that tend to favour $A_u<0$ if
$\mu>0$ and therefore rule out large, positive values of $A_u$ in both
plots.

\section{Constraints on Multiple Insertions}

\begin{figure}[!p]
\begin{center}
\begin{tabular}{c c}
	\includegraphics[width=0.00\textwidth]{spacer.eps}
	\includegraphics[width=0.40\textwidth]{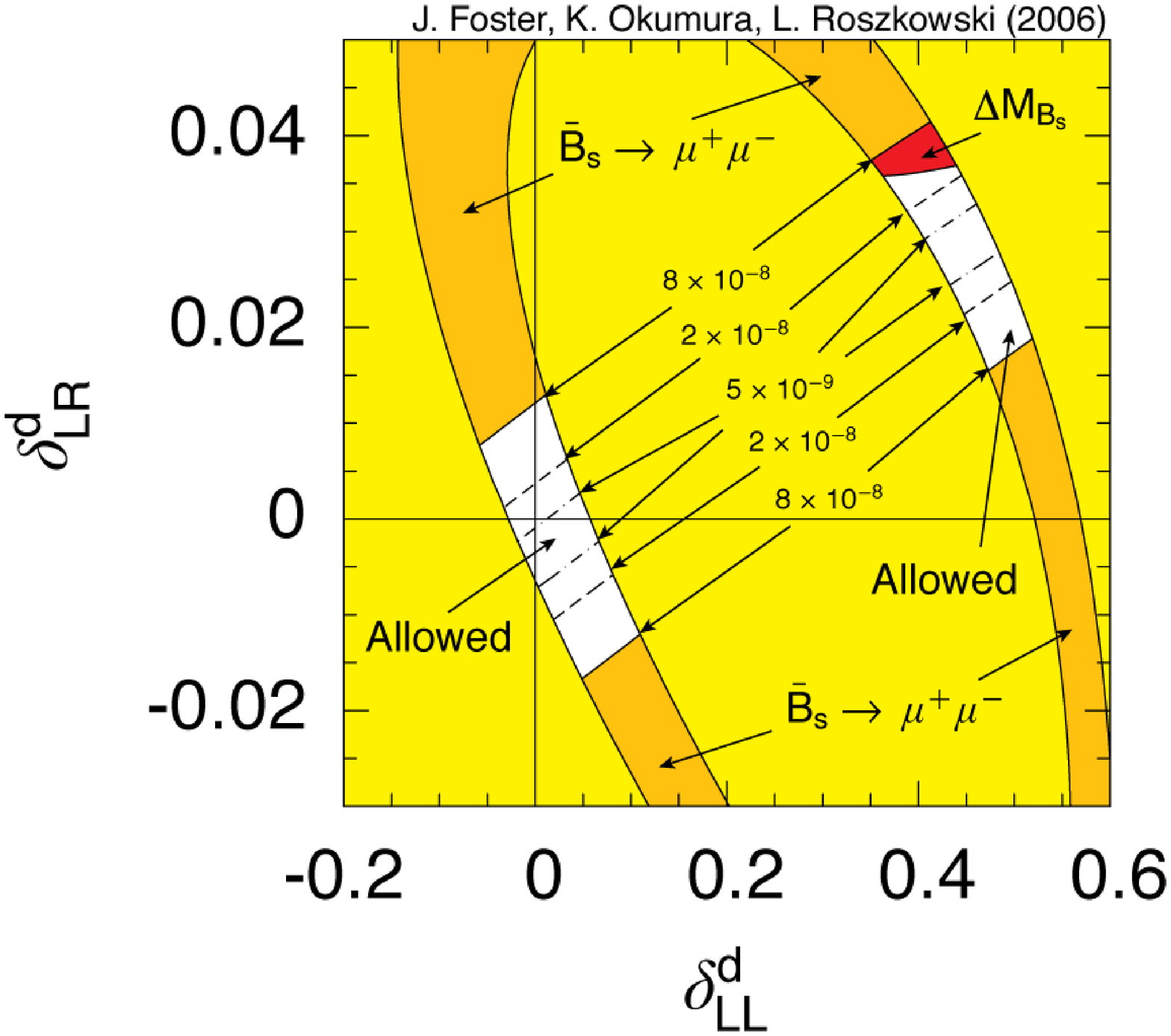}
	\includegraphics[width=0.00\textwidth]{spacer.eps}
&	
	\includegraphics[width=0.02\textwidth]{spacer.eps}
	\includegraphics[width=0.435\textwidth]{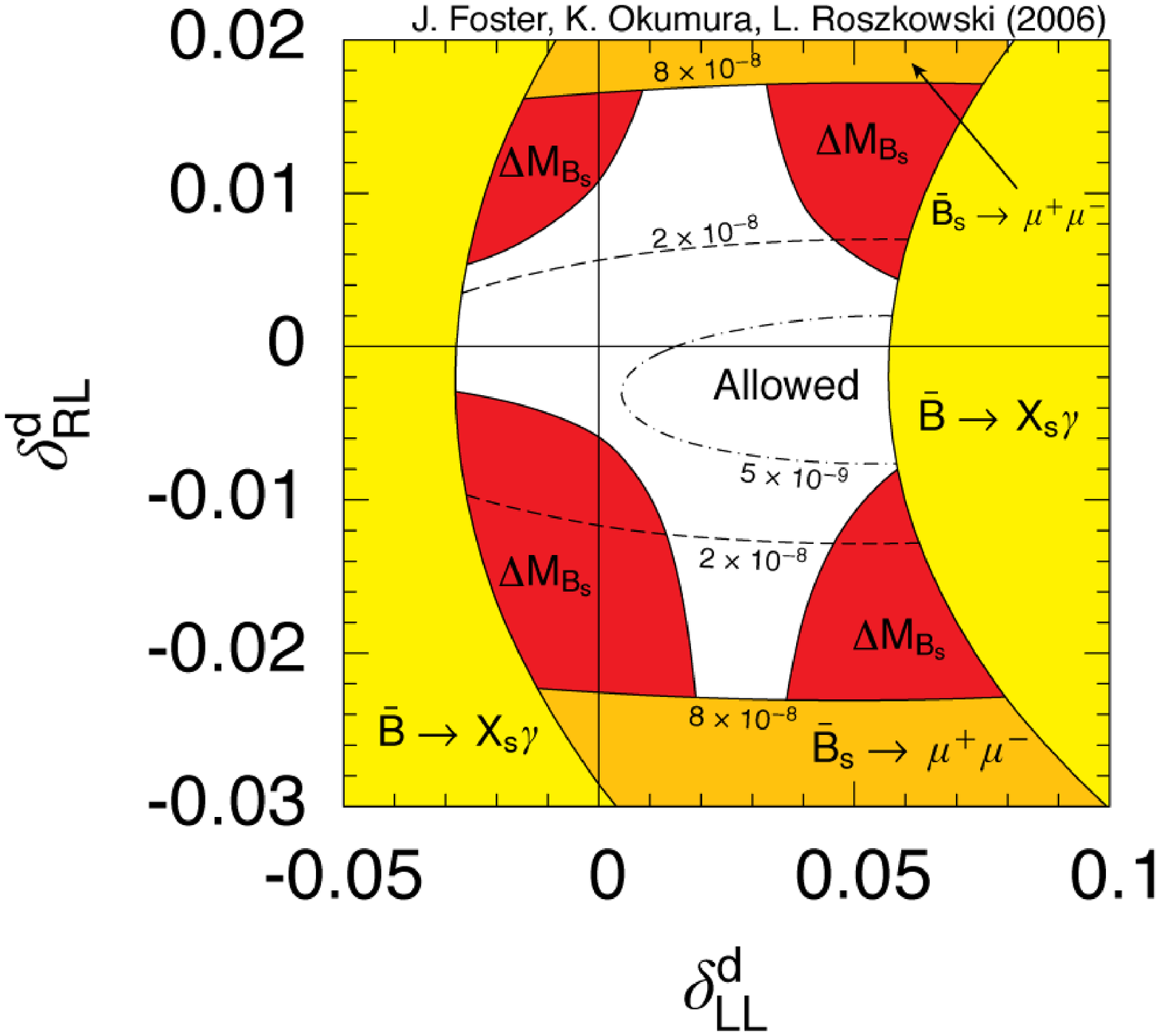}
	\includegraphics[width=0.00\textwidth]{spacer.eps}
\\
	\includegraphics[width=0.04\textwidth]{spacer.eps}
	\includegraphics[width=0.425\textwidth]{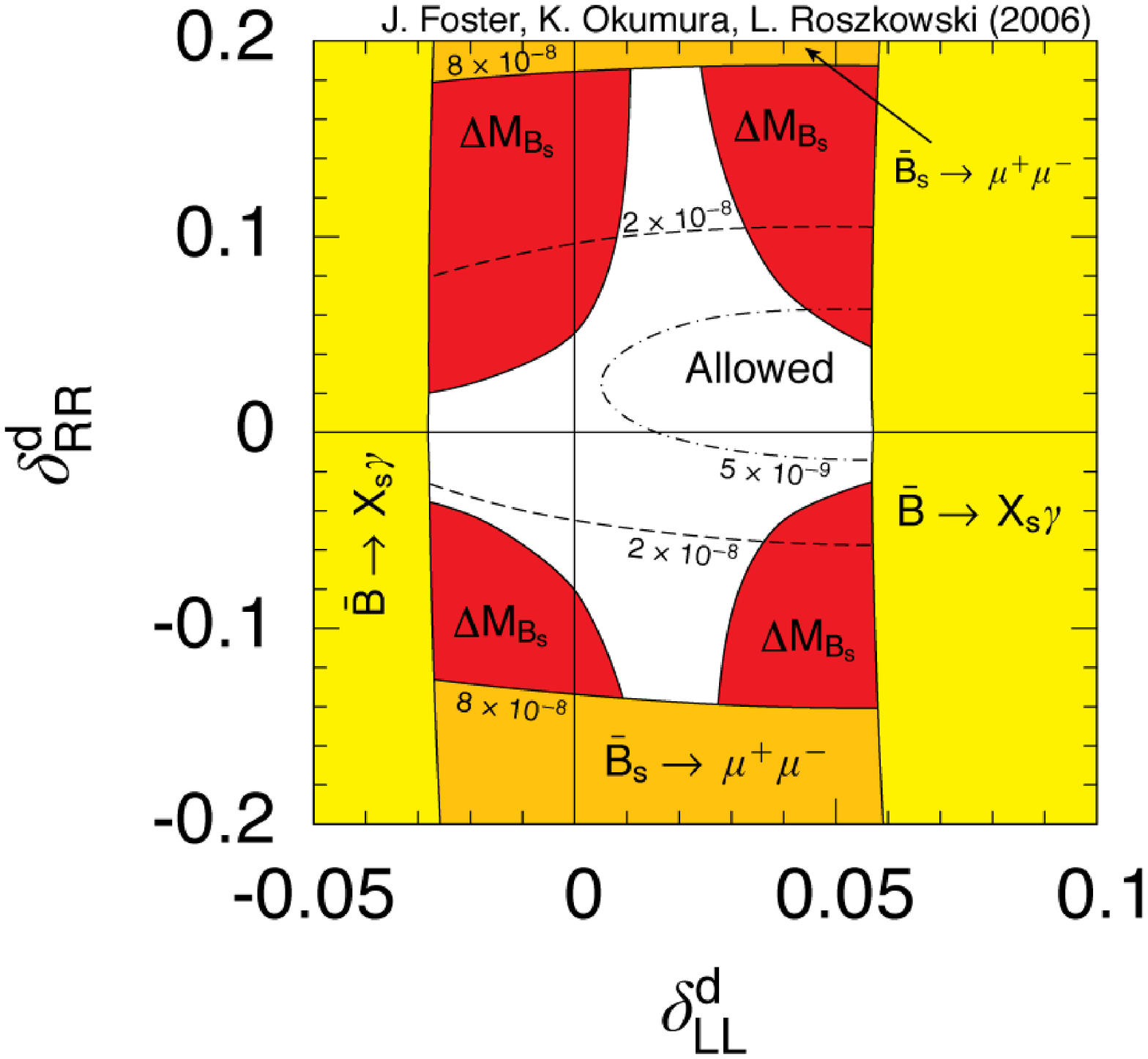}
	\includegraphics[width=0.00\textwidth]{spacer.eps}
&	
	\includegraphics[width=0.00\textwidth]{spacer.eps}
	\includegraphics[width=0.42\textwidth]{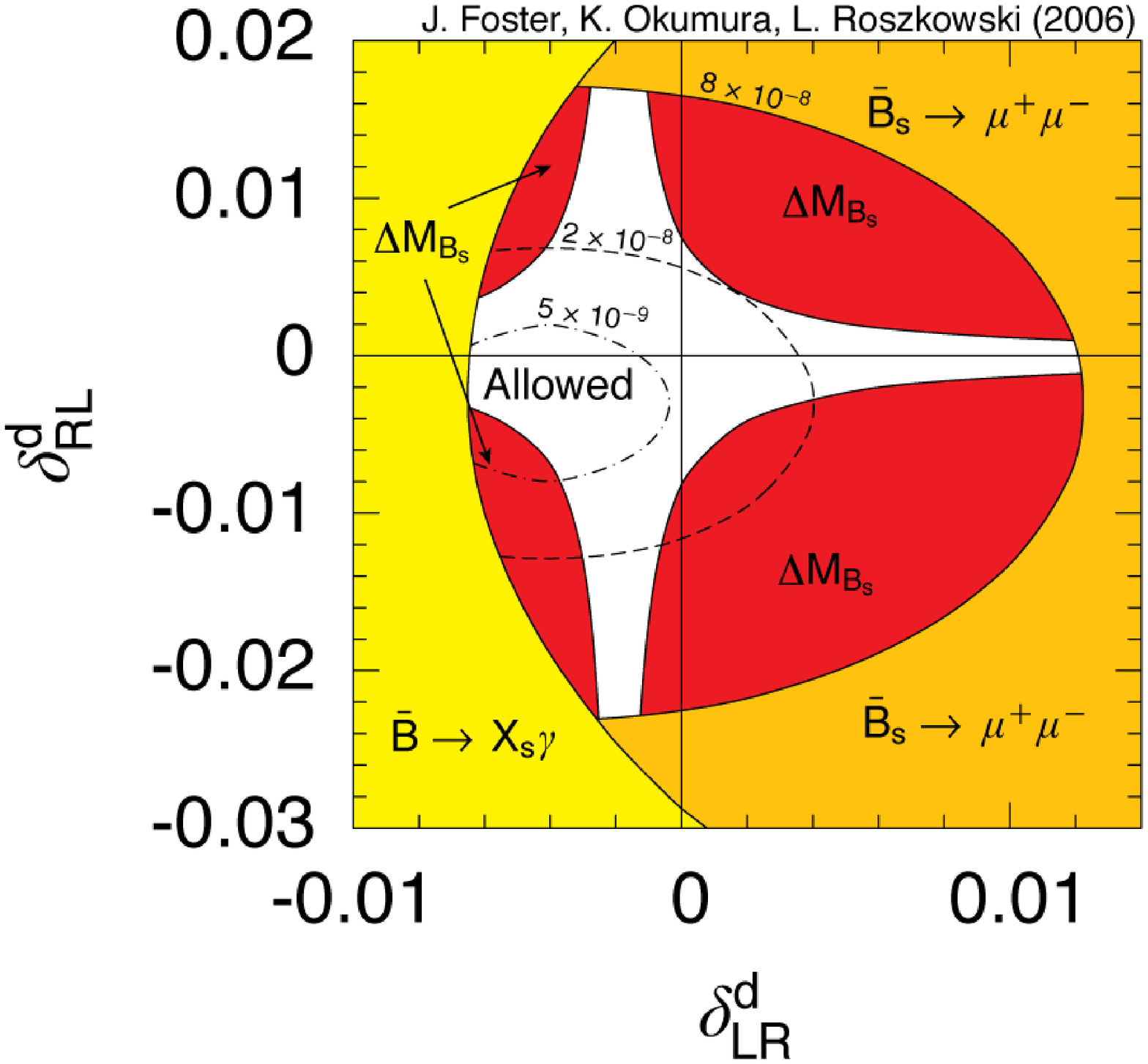}
	\includegraphics[width=0.00\textwidth]{spacer.eps}
\\
	\includegraphics[width=0.015\textwidth]{spacer.eps}
	\includegraphics[width=0.42\textwidth]{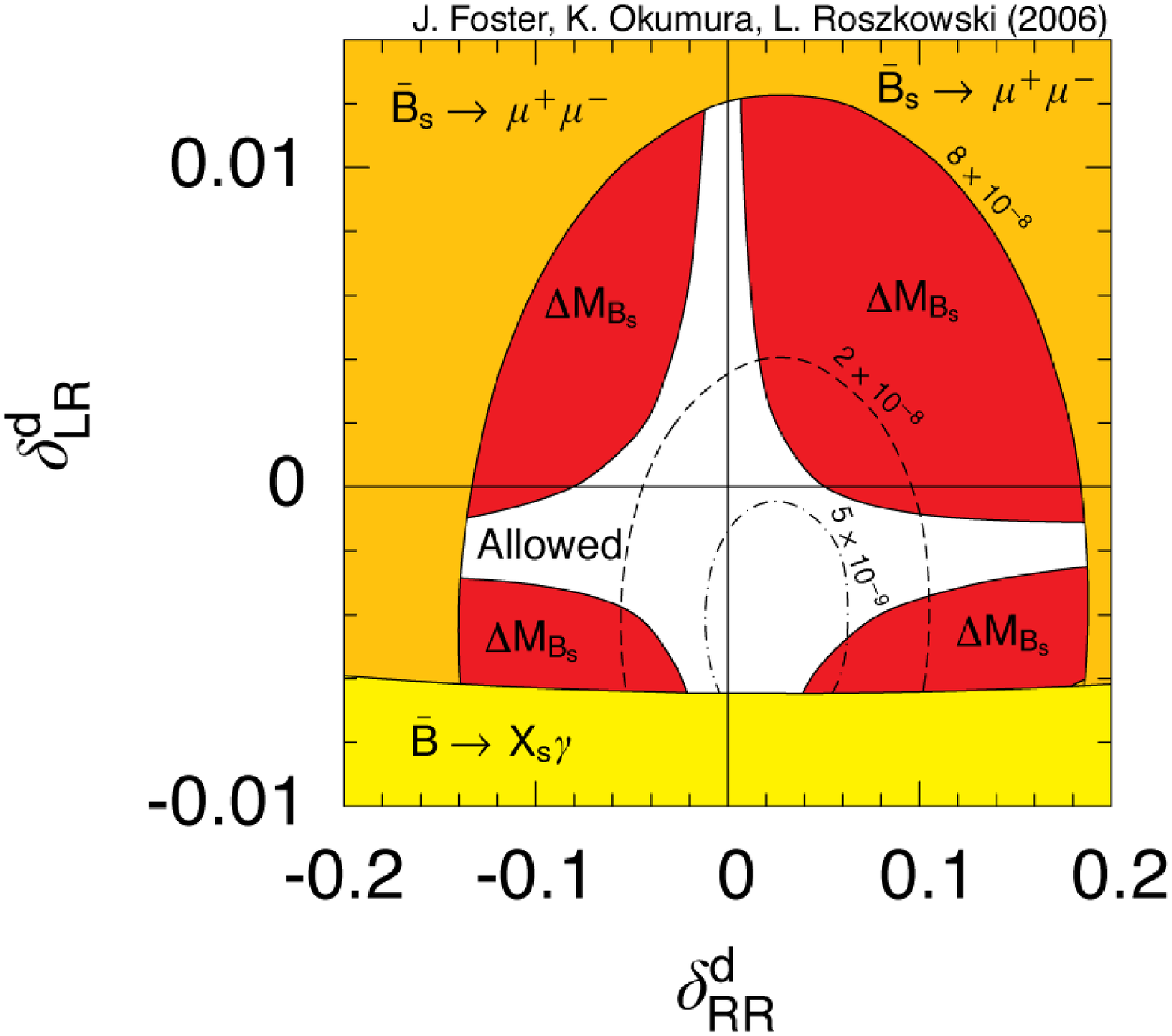}
	\includegraphics[width=0.00\textwidth]{spacer.eps}
&	
	\includegraphics[width=0.00\textwidth]{spacer.eps}
	\includegraphics[width=0.43\textwidth]{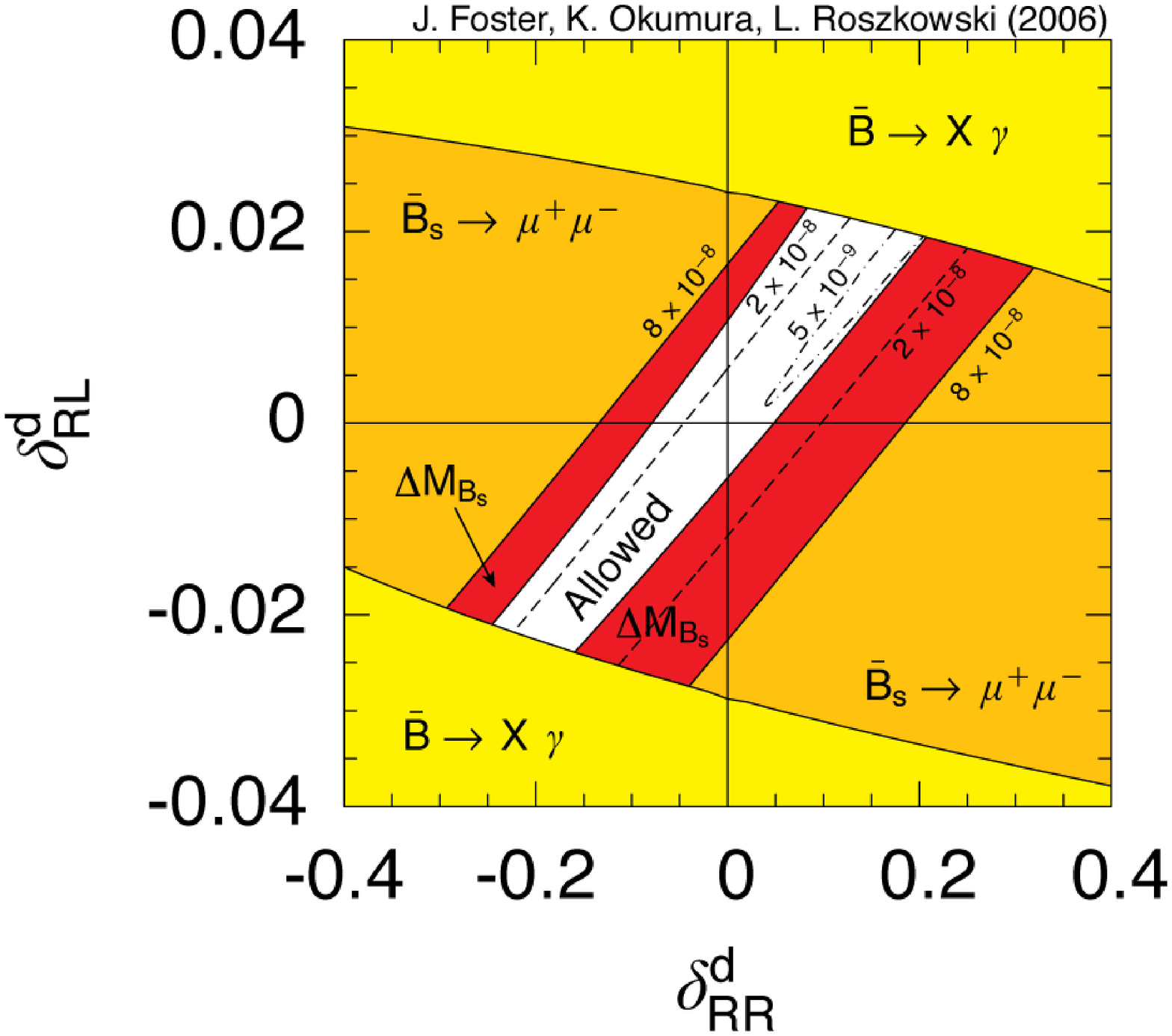}
	\includegraphics[width=0.00\textwidth]{spacer.eps}
\end{tabular}
\end{center}
\caption{The soft sector is parameterised as follows $\msq=\mgl=1\tev$
  and $m_A=\mu=-A_u=500\gev$ and $\tanb=40\gev$. The excluded and
  allowed regions are shaded in a similar manner
  to~\fig{FIG1}.\label{FIG5} }
\end{figure}
As $B_s$ mixing is a $\Delta F=2$ process it is naturally more sensitive
to scenarios where more than one insertion is present at a given time.
It is therefore useful to consider how the new results
regarding $B_s$ mixing will affect the bounds on multiple sources of flavour
violation. If one refers to the formula given in~\eqref{EQ1} 
in the previous section, it can be seen that exceptionally large
contributions to $B_s$ mixing are possible when an LL or an LR insertion
is varied at the same time as a RL or RR insertion. These large contributions
to $B_s$ mixing can easily be in conflict with current limits provided
by D\O~\eqref{BBexpd0} and CDF~\eqref{BBexpcdf} and they will, therefore, provide an
excellent constraint 
in such scenarios. Such a situation is illustrated in~\fig{FIG5}
where we show the six possible combinations that can be formed when
varying two insertions at a time.

From the figure it can be seen immediately that the new bound imposed
by $B_s$ mixing has a substantial effect on the available parameter
space in all but one of the six panels. (The $\dll$--$\dlr$ parameter
space is left relatively unaffected by the $B_s$ mixing constraint as
the corresponding contribution to $\delmbs$ is suppressed by
$m_s/m_b$.)  In fact, in the four panels constraining the combination
of an LL or LR insertion together with an RL or RR insertion it can be
seen that the constraints on the available parameter space are almost
completely dominated by $B_s$ mixing with the additional constraints
supplied by either $\bsg$ or $\bsm$ only ruling out regions where either RL
or RR insertions are small, or fine tuned regions where an LL or LR
insertion accidentally cancels with the MFV contribution to the
neutral Higgs penguin.

Also illustrated in all six panels is the effect that an improvement
in the determination of the upper bound on the branching ratio
for $\bsm$ might have. From all of the panels it is evident
that improving the limit to $2\times 10^{-8}$ will further reduce the
available parameter space. However, in all but the top left panel
it is also apparent 
that the lower and upper bounds on $B_s$ mixing play a far more important
role in constraining certain combinations of the insertions. Indeed, in
much of the parameter space,
only after the limit on $\bsm$ has reached a level approaching
$5\times 10^{-9}$ would it provide a bound exceeding that imposed
by $B_s$ mixing when constraining multiple pairs of insertions.

\section{Conclusions}

In this Letter we have summarised the new constraints that can be
placed on SUSY flavour violation in light of the recent limits placed
on $\delmbs$ and the improved bound on $\bsm$ at the Tevatron. While
unfortunately this observation has not signalled the presence of
physics beyond the Standard Model, it now places strict new
constraints on SUSY flavour violation. In particular, RR and RL
insertions are now constrained to a similar degree as LL and LR
insertions in the large $\tanb$ regime. Consequently models that
predict large values for these insertions (such as grand unified
theories that incorporate a SUSY seesaw model, for
instance~\cite{GUT}) might now encounter strict constraints in the
large $\tanb$ regime unless the Higgs sector is naturally very heavy
(\ie~$m_A\gg 1\tev$). In addition, this observation seems to rule out
large contributions to $\bsm$ arising from RR insertions and any
forthcoming measurement of $\bsm$ at the Tevatron would be hard to
reconcile with the effect of RR insertions unless $\vert A_u\vert$ is small.
\\
\\ 

{\bf Acknowledgements} \\ 
We would like to thank A.~Masiero for
helpful comments. This work was partially supported by the EC 6th
Framework Programme MRTN-CT-2004-503369.  J.F. has been supported by
the research fellowship MIUR PRIN 2004 -- ``Astroparticle
Physics''. K.O. has been supported by the grant-in-aid for scientific
research on priority areas (No. 441): ``Progress in elementary
particle physics of the 21st century through discoveries of Higgs
boson and supersymmetry'' (No. 16081209) from the Ministry of
Education, Culture, Sports, Science and Technology of Japan and the
Korean government grant KRF PBRG 2002-070-C00022.  L.R. would like to
thank A.~Masiero  and INFN Padova for kind hospitality during his recent visit when
this work was finalised.

\begin{appendix}

\section*{Appendix}

The factors of $\epsilon$ that enter into~\eq{EQ1} have the form~\cite{FOR}
\begin{align}
\epsg=&-\frac{\alpha_s}{2\pi}C_2(3)\frac{\mu}{\mgl}
H_2\left(\xdr,\xdl\right), & 
\epsy=&-\frac{A_u}{16\pi^2\mu}
H_2\left(\yur,\yul\right),
\nonumber\\
\epsilon_{RL}=&-\frac{\alpha_s}{2\pi}C_2\left(3\right)
H_2\left(\xdr,\xdl\right), &
\epsilon_{LL}=&-\frac{\alpha_s}{2\pi}C_2(3)\frac{\mu}{\mgl}
H_3\left(\xdr,\xdr,\xdl\right),
\nonumber\\
\epsilon_3=&\epsg+\epsy Y_t^2
\nonumber
\end{align}
where $\alpha_s$ denotes the strong coupling constant, $C_2(3)=4/3$
is the quadratic Casimir for $\suthree$ and $A_u$ is related to the
soft terms $m^2_{u,LR}$ via the relation $\left(m^2_{u,LR}\right)=A_u m_u$.
The loop functions that appear in the above expressions are
\begin{align}
H_2(x_1,x_2)=&\frac{x_1\log x_1}{\left(1-x_1\right)\left(x_1-x_2\right)}
+\frac{x_2\log x_2}{\left(1-x_2\right)\left(x_2-x_1\right)},
\nonumber\\
H_3(x_1,x_2,x_3)=&\frac{H_2(x_1,x_2)-H_2(x_1,x_3)}{x_2-x_3}.
\nonumber
\end{align}
Finally, the arguments of the various 
loop functions are given by
\begin{align}
\xdl&=\frac{m_{d,LL}^2}{\mgl^2},
&\yul&=\frac{m_{u,LL}^2}{\mu^2},
&\xdrl&=\frac{\sqrt{m_{d,LL}^2 m_{d,RR}^2}}{\mgl^2}.
\nonumber
\end{align}
The remaining factors of $\epsilon$, $x$ and $y$ may be obtained by
the appropriate generalisations of the above expressions. Note that
the soft terms that appear in the above equation are the common values
of the diagonal entries of the SUSY soft terms.

\end{appendix}


\end{document}